\documentclass{aa}
\usepackage{graphicx}
\usepackage{txfonts}
\begin{document}
    \title{Hyades dynamics from N-body simulations:\\
Accuracy of astrometric radial velocities from Hipparcos
\thanks{Based on observations by the ESA Hipparcos satellite,
and on the $N$-body code NBODY6 by Sverre Aarseth, publicly
available at ftp://ftp.ast.cam.ac.uk/pub/sverre/}
}
    \author{S{\o}ren Madsen}
    \offprints{S. Madsen, {\tt soren@astro.lu.se}}
    \institute{
    Lund Observatory, Box~43,
    SE--22100 Lund, Sweden\\
    \email{soren@astro.lu.se}}

\date{Received 30 October 2002 / Accepted 27 January 2003}
\titlerunning{Hyades dynamics from N-body simulations}
\authorrunning{S{\o}ren Madsen}

\abstract{
The internal velocity structure in the Hyades cluster as seen by Hipparcos
is compared with realistic $N$-body simulations using the NBODY6 code,
which includes binary interaction, stellar evolution and the Galactic
tidal field. 
The model allows to estimate reliably the accuracy of astrometric radial velocities
in the Hyades as derived by Lindegren et al.\ (\cite{lindegren00}) and
Madsen et al.\ (\cite{madsen02}) from Hipparcos data, by applying the same
estimation procedure on the simulated data. The simulations indicate that
the current cluster velocity dispersion decreases from
0.35~km~s$^{-1}$ at the cluster centre to a minimum of 0.20~km~s$^{-1}$
at 8~pc radius (2--3 core radii), from where it slightly increases outwards. A clear
negative correlation between dispersion and stellar mass is seen in the central
part of the cluster but is almost absent beyond a radius of 3~pc. 
It follows that the (internal) standard error of
the astrometric radial velocities relative to the cluster centroid may be as
small as 0.2~km~s$^{-1}$ for a suitable selection of stars, while a total
(external) standard error of 0.6~km~s$^{-1}$ is found when the uncertainty
of the bulk motion of the cluster is included. 
Attempts to see structure in the velocity dispersion using observational data
from Hipparcos and Tycho-2 are inconclusive.
\keywords{
Methods: N-body simulations -- data analysis --
Techniques: Astrometry --
Stars: kinematics --
Open clusters and associations: general --
Open clusters and associations: individual: Hyades}}

\maketitle


\section{Introduction}

The Hyades is the nearest rich open cluster and as such has played a fundamental
role in astronomy as a first step on the cosmic distance ladder and as a test case
for theoretical models of stellar interiors (Lebreton \cite{lebreton00}). From the
first use of the converging point method by Boss (\cite{boss08}) up to the use
of pre-Hipparcos trigonometric parallaxes by van Altena et al.\ (\cite{altena97}),
an important goal in astrometry has been the determination of an accurate distance
to the cluster. With the advent of the Hipparcos Catalogue (ESA \cite{esa}) the Hyades
lost its unique status for distance calibration, but as the depth and internal
velocity field of the cluster were well resolved by Hipparcos, focus could
instead be turned to its three-dimensional structure and kinematics
(Perryman et al.\ \cite{perryman98}). A deeper understanding of the dynamics and
evolution of the cluster should now be possible through detailed comparison with
$N$-body simulations.

Thanks to the accurate Hipparcos measurements, the Hyades has recently acquired
a completely new role as a practical standard in observational astrophysics: it
is one of very few objects outside the solar system for which the accurate radial
motion can be determined by geometric means, i.e.\ without using the
spectroscopic
Doppler effect. From a combination of Hipparcos parallaxes and proper motions,
Madsen et al.\ (\cite{madsen02}) obtained ``astrometric radial velocities" for
individual Hyades stars with a then estimated standard error of about 0.6~km~s$^{-1}$.
Currently
the Hyades is the only cluster for which astrometric radial velocities are
derived
with individual accuracies better than 1~km~s$^{-1}$, but the technique may be
extended to many more objects with future space astrometry missions (Dravins
et al.\ \cite{dravins99b}).

Astrometric radial velocities are important mainly because they make it possible
to determine the {\em absolute} lineshifts intrinsic to the stars, through
comparison with spectroscopic measurements.  Such lineshifts are caused for
instance by convective motions and gravitational redshift in the stellar
atmospheres (Dravins et al. \cite{dravins99a}). Absolute lineshifts could
previously only be observed
in the solar spectrum, but are now within reach for a range of spectral types
through the use of astrometric radial velocities. The present paper is part
of a research programme at Lund Observatory in which absolute lineshifts are
determined and used as a diagnostic tool in stellar astrophysics
(Dravins et al.\ \cite{dravins97}, \cite{dravins99b};
Lindegren et al.\ \cite{lindegren00}; Madsen et al.\ \cite{madsen02};
Gullberg \& Lindegren \cite{gullberg02}).

A major uncertainty in the astrometric radial velocities originates in the
internal velocity dispersion of the cluster, which limits both the accuracy
of the cluster motion as a whole, and that of the individual stars.
A primary goal of the present investigation is to find out whether a better
understanding of the internal velocity structure of the cluster, obtained
through $N$-body calculations, can be used to improve the accuracy of the
astrometric radial velocities.

Sect.~\ref{sec:kin} briefly recalls the kinematic information, including
astrometric radial velocities, that can be derived from Hipparcos data.
Sect.~\ref{sec:dyn} describes the model used to simulate the evolution of
the cluster up to its present state, and its subsequent observation, as
well as the main properties derived from the simulations. Implications for
the accuracy of the astrometric radial velocities are discussed in
Sect.~\ref{sec:pred}, followed by a discussion of non-modelled effects
in Sect.~\ref{sec:nmod}, and conclusions.

\section{Cluster kinematics derived from astrometry}
\label{sec:kin}

Since an ultimate aim of the present programme is to confront spectroscopic
measurements of line shifts in stellar spectra with independent measurements
of the stellar motions, it is essential that the kinematic data, including the
radial velocities, are derived without using the spectroscopic Doppler effect.
Dravins et al.\ (\cite{dravins99b}) describe several methods to derive the
radial motion of stars by purely geometric means, i.e.\ using astrometric data.
Of these, the moving-cluster method has been successfully applied to several
open clusters and OB associations, in particular the Hyades (Lindegren et al.\
\cite{lindegren00}; Madsen et al.\ \cite{madsen02}). The principle of the
moving-cluster method is very simple: let $\theta$ be the angular size of the
cluster and $R$ its distance. Assuming its linear size $R\theta$ to be constant,
we have $\dot{R}\theta+R\dot{\theta}=0$, where the dot signifies time
derivative. Since $R$ is known from trigonometric parallaxes, the astrometric
radial velocity of the cluster follows as $\dot{R}=-R\dot{\theta}/\theta$.

In practice, several kinematic parameters are simultaneously estimated from
the astrometric data of the cluster member stars, using the method of maximum
likelihood (Lindegren et al.\ \cite{lindegren00}). Some features of the method,
relevant for the subsequent discussion, are recalled hereafter.

The estimated parameters include the common space velocity of the cluster
($\vec{v}_0$), the individual stellar parallaxes ($\pi_i$ for star
$i$), and the internal velocity dispersion ($\sigma_v$). The astrometric
radial velocity of an individual star $i$ is then calculated as
$\widehat{v}_{{\rm r}i}=\vec{r}_i'\widehat{\vec{v}}_0$, where $\vec{r}_i$
is the unit vector towards the star and the caret $\widehat{\phantom{x}}$
signifies estimated quantities. As part of the procedure, improved
parallaxes $\widehat{\pi}_i$ are obtained for the individual stars.
In the Hyades, these are 2--5 times more precise than the original
Hipparcos parallaxes which have errors around 1--1.5 mas. 
The improvement results from a combination of
trigonometric and kinematic parallaxes, where the latter follow from
the proper-motion components \emph{along} the cluster motion, which are
inversely proportional to distance. The kinematically improved
parallaxes allow a very precise mapping of the spatial structure of
the cluster. The maximum likelihood estimate of $\sigma_v$ is 
unfortunately biased. Instead the proper motions \emph{perpendicular} to the cluster
motion are used to estimate the velocity dispersion according to the method described in Lindegren
et al.\ (\cite{lindegren00}), Appendix~A.4. For each star,
a goodness-of-fit statistic $g_i$ is also obtained from the
maximum-likelihood estimation (see Lindegren et al.\ (\cite{lindegren00})
for a thorough discussion of $g_i$).
The statistic is primarily used to reject stars
whose astrometric data do not fit the cluster model well enough;
a rejection limit of $g_{\rm lim}=15$ was normally used, although
a stricter limit (10) or no limit at all ($\infty$) were also tried.
For the retained stars, the $g_i$ values (which are then
$\le g_{\rm lim}$) could be regarded as a quality index, with a lower
value indicating a better fit to the cluster model.

The error in the estimated astrometric radial velocity,
$\widehat{v}_{{\rm r}i}$,
has two parts. The first part is due to the error in the common space motion
of the cluster, $\widehat{\vec{v}}_0$. Its uncertainty depends on global
properties of the cluster such as its distance, angular extent, and richness,
as well as on the accuracy of the astrometric data. The second part is due to
the star's peculiar motion relative to the cluster centroid.
This part depends only on the dispersion of the
peculiar motions along the line of sight, which for a uniform, isotropic
velocity dispersion equals $\sigma_v$. In most of the clusters for which the
method has been applied, the main uncertainty comes from the first part,
i.e.\ the error in the cluster's space motion. In the Hyades, however, the
uncertainty in $\widehat{\vec{v}}_0$ is small enough (0.36~km~s$^{-1}$
along the line-of-sight; Madsen et al.\ \cite{madsen02}) that the total
uncertainty in the astrometric radial velocities is dominated by the
contribution from the internal velocity dispersion (0.49~km~s$^{-1}$
according to the estimate in the same source).

On the other hand, the assumption of a constant and isotropic velocity
dispersion throughout the cluster may be rather simplistic. Theoretically,
one expects at least a variation with distance $r$ from the centre of the
cluster, and possibly also a variation with stellar mass due to the
equipartition of kinetic energy. For instance, in a simple Plummer (\cite{plummer15})
potential we have
\begin{equation}
    \label{eq:plum}
    \sigma_v^2(r) = \frac{GM}{6\sqrt{r_c^2+r^2}}
\end{equation}
(Gunn et al.\ \cite{gunn88}; Spitzer \cite{spitzer87}),
where $M$ is the cluster mass and $r_c$ the core radius
($\simeq 3$~pc for the Hyades). According to Eq.~(\ref{eq:plum}),
$\sigma_v$ should decrease by one third as one moves two core radii
away from the centre, and become even smaller further out in the cluster;
but this trend is obviously broken at some distance by tidal forces.
Clearly, these effects must be also reflected in the accuracy of the
astrometric radial velocities. Attempts to measure the radial variation
of dispersion in the Hyades from astrometry 
were inconclusive (Madsen et al.\ \cite{madsen01}), but
$N$-body simulations could help to establish to what extent such
variations exist in real clusters.

\section{Dynamical simulation of the Hyades cluster}
\label{sec:dyn}

\subsection{Previous N-body simulations}
\label{sec:prev}

It is not new to use the Hyades as a comparison with $N$-body simulations.
Aarseth (\cite{aarseth77}) discussed the dynamical relevance of the central
binary 80~Tau (HIP~20995) in the context of binary formation
and evolution in stellar systems as described by $N$-body simulations.
Oort (\cite{oort79}) discussed
the flattening of the Hyades parallel to the galactic plane by comparing
observations with the $N$-body simulations by Aarseth (\cite{aarseth73}).
Kroupa (\cite{kroupa95c}) simulated the evolution of star clusters and
found excellent agreement between the models and the Hyades luminosity
function, concluding that the initial conditions of
the cluster could to a large extent be reconstructed. An initial mass of
the Hyades protocluster of some $1300~M_{\odot}$ was suggested.
von Hippel (\cite{hippel98}) used numerical simulations of clusters and
data on Hyades white dwarfs, among others, to conclude that the white-dwarf
mass fraction is relatively insensitive to kinematic evolution.
Portegies Zwart et al.\ (\cite{portegies01}) discussed the evolution of
star clusters which were given initial conditions to represent open clusters,
including the Hyades. A good model fit to the Hyades was obtained, thus
illustrating the possibility to estimate the initial conditions
for an observed star cluster.

What is new in the present study is that the three-dimensional kinematics of
the Hyades is investigated through a direct comparison of the Hipparcos
observations with a realistic $N$-body model, evolved till the present age
of the cluster, as well as the objective to estimate the accuracy of the
astrometric radial velocities from such a comparison.

\subsection{Basic cluster data}
\label{sec:bas}

Perryman et al.\ (\cite{perryman98}) made a detailed study on the Hyades based
on Hipparcos data and a compilation of spectroscopic radial velocities from
the literature. They identified 197 probable member stars, which constitute
the initial Hyades sample (Hy0) used for the present study. 
When comparing with the simulated cluster,
only stars within 20~pc from the cluster centre are considered, due to the
radial limitation in the $N$-body code (Sect.~\ref{sec:mod}). Adopting
the cluster centre of mass in equatorial coordinates, $(+17.36,+40.87,+13.30)$~pc
from Perryman et al.\ (\cite{perryman98}), and using the kinematically improved
parallaxes (Sect.~\ref{sec:kin}), a subset of 178 stars (Hy0r) was found
within a radius of 20~pc. The cluster has a general space velocity of
$(-5.90,+45.65,+5.56)$~km~s$^{-1}$ in equatorial coordinates (Madsen et al.\ \cite{madsen02}).

Perryman et al.\ (\cite{perryman98}) note that
a redetermination of membership with the above cited centre of mass will reduce
the number of member stars outside 10 pc by 10 stars while keeping
the same number of stars inside 10 pc. The true number of member stars
in the Hy0r sample is then probably smaller than the 178 stars.

Hipparcos is nominally complete to $V \le 7.3 + 1.1|\sin b|$ for spectral
types later than G5 (or $B\!-\!V>0.8$). However, it is known that the actual
limit is somewhat fuzzy, due to photometric errors and other complications.
Therefore, a conservative completeness limit of $V \le 7$~mag is assumed for
this study. 
Choosing a fainter completeness limit like e.g. $V \le 8$~mag will, however, not 
significantly affect the outcome of the simulations as it will be shown later (Table~1).
The actual number of fainter Hyades members is not known.
However, at least seven single white dwarfs have been found
(e.g.\ Reid \cite{reid96}), and this number can also be used as a constraint
on the model.

Perryman et al.\ (\cite{perryman98}) estimated the cluster age to be
$625 \pm 50$~Myr, and this age is what is assumed in the following.
It should be mentioned that in a more recent work by Lebreton et al.\ (\cite{lebreton01}),
based on kinematically improved parallaxes from Dravins et al.\
(\cite{dravins97}), only an upper limit of 650~Myr could be estimated due to
the lack of a clear turn-off point (cf.\ top diagram in Fig.~\ref{fig:cmd}).
In the same work they
also estimated the metallicity to $\mbox{[Fe/H]}=0.14 \pm 0.05$~dex.
The interstellar extinction is negligible:
Taylor (\cite{taylor80}) found only a
very small colour excess $E(B-V)=0.003 \pm 0.002$~mag.

From various studies, a large fraction of the stars are known to be binaries.
In the compilation by Perryman et al.\ (\cite{perryman98}), 75 of the 197
probable member stars were either identified as binaries in the Hipparcos
Catalogue or previously known as spectroscopic binaries (their Table~2).
Patience et al.\ (\cite{patience98}) found three new binaries from a speckle
imaging survey of Hyades members, plus one marked as binary in the Hipparcos
Input Catalogue (HIC; Turon et al.\ \cite{turon92}), but not found by Hipparcos.
In the Tycho Double Star Catalogue (Fabricius et al. \cite{fabricius02}), an additional
21 binaries were identified. The eclipsing binary system HIP~17962 = V471~Tau
(e.g.\ Werner \& Rauch \cite{werner97}, and references therein) must
also be included in the list of Hyades binaries.
We thus end up with 101 known binaries
in the Hy0 sample, yielding a minimum multiplicity of 0.51 companions per
primary. For the Hy0r sample (within 20~pc of the cluster centre) the minimum
multiplicity is 0.53. To include some more binary statistics, binaries with periods $P < 10$ days
have been taken from the compilations on the open--cluster database 
WEBDA\footnote{available at http://obswww.unige.ch/webda/}.

The above values of the multiplicity are only lower limits to the true multiplicity, because of
the difficulty to detect binaries in some intervals of separation $\rho$ (or period $P$)
and magnitude difference $\Delta m$ (or mass ratio $q$). In restricted intervals, the
searches can however be considered complete. For instance, Hipparcos probably
detected practically all binaries with $0.2 < \rho < 2$~arcsec and $\Delta m < 2$; cf.\
Fig.~3.2.106 in Vol.~1 of (ESA \cite{esa}), where 17 are found in Hy0r.
Patience et al.\ (\cite{patience98}) observed a high fraction
of Hyades stars that were also observed by Hipparcos. The 17 binaries
they found with $0.1 < \rho < 1.07$~arcsec and $q\ge 0.4$ must therefore also
be regarded as a nearly complete sample.

Hipparcos effectively observed for about 37 months ($\sim$3~years) spread
over a period of nearly 4~years. This means that the proper
motions of binaries may be significantly affected by the orbital motion
of the photocentre, which must be taken into account when simulating the
Hyades proper motions (Sect.~\ref{sec:tran}). In order to reduce this effect
in the observational analysis, proper motions from the Tycho-2 catalogue
(H{\o}g et al.\ \cite{hog00}) have also been used, where available.
In the solution for the cluster kinematics, the Tycho-2 proper motions
yield slightly, but systematically smaller radial velocities
($v_r($HIP$)-v_r($Tycho-2$)=+$0.9~km~s$^{-1}$) than do the Hipparcos data
for the $g_{\rm lim}=15$ sample   (Madsen et al.\ \cite{madsen02}), which can be
explained by the mean difference of -0.4~mas~yr$^{-1}$ of the proper motions
in right ascension of what was considered the best sample. In declination,
the mean difference of the proper motions is 0.0~mas~yr$^{-1}$.
Although the expected deviations between the Hipparcos and Tycho-2 Catalogues are
generally under 0.5~mas (Urban et al. \cite{urban00}), the result from
the Hyades might reflect some subtle bias in the Tycho-2 proper-motion system.
Since the Tycho-2 system of proper motions was effectively calibrated
onto the Hipparcos system, greater confidence should be put on the
solution based on the Hipparcos data.
The Tycho-2 data should therefore only be used to study the internal
velocity structure of the cluster, where a possible bias is not important.

In addition to the Hy0r sample (which thus includes all 178 probable members
within a radius of 20~pc from the cluster centroid), the following samples
are also discussed: Ty0r, which is the same as Hy0r but with proper motions
from Tycho-2 replacing those in the Hipparcos Catalogue; Hy1r, which is the
subset of 85 stars in Hy0r for which there is no known indication of
multiplicity; and Ty1r, which is the same as Hy1r but with Tycho-2 proper
motions.

It has been suggested that there might be systematic errors in the Hipparcos
parallaxes for at least some open clusters (Pinsonneault et al. \cite{pins98}).
The discussion shall not be repeated here, but it should just be stated
that there is a general consensus that the mean Hyades parallax is not affected
by any correlation between positions and parallaxes (Narayanan \& Gould \cite{narayanan99};
van Leeuwen \cite{vanleeuwen00}; Lindegren et al. \cite{lindegren00}).
This problem, if it exists, has been neglected in the simulations.

\subsection{N-body model of the Hyades cluster}
\label{sec:mod}

The dynamical evolution of a Hyades-type open star cluster was simulated
using the well-known $N$-body code NBODY6 (Aarseth \cite{aarseth99},
\cite{aarseth00}). The code incorporates algorithms to deal with stellar
(including binary) encounters (Mikkola \& Aarseth \cite{mik93}, \cite{mik96},
\cite{mik98}) and stellar evolution (Hurley et al.\ \cite{hurley00}).
For the present study, no modifications were made to the code. Some of the
non-modelled effects are discussed in Sect.~\ref{sec:nmod}.

External perturbations are represented by a fixed, galactic tidal field.
The cluster is assumed to move in a circular orbit at the present distance
of the Sun from the galactic centre. The angular velocity is $\Omega=A-B$,
where $A=14.4$~km~s$^{-1}$~kpc$^{-1}$ and $B=-12.0$~km~s$^{-1}$~kpc$^{-1}$
are Oort's constants. This gives rise to tidal forces 
plus a Coriolis force (cf.\ Chandrasekhar \cite{chandra42}, Ch.~5.5).

To set up the initial cluster configuration, stars are randomly picked from
the initial mass function (IMF) described by Kroupa et al.\ (\cite{kroupa93}),
until the required total particle number has been reached. Binaries are included
as described below. Stars are initially deployed randomly in a Plummer potential
(Plummer \cite{plummer15}; Spitzer \cite{spitzer87}) with virial radius
$r_v=4$~pc. During the evolution
of the cluster, stars are kept in the simulation as long as they are within
two tidal radii ($\simeq 21$--$23$~pc). The simulation is run until the cluster
reaches an age of 625~Myr.

The reason for choosing one single age was to have a fixed parameter
for comparing different model realisations. The age uncertainty
is not important regarding the conclusions about the
current dynamics since the cluster has been relaxed for quite a
while.

Binaries are generated by randomly pairing stars picked from the IMF.
This gives an almost uniform distribution in the logarithm of the
mass ratio ($\log q$), i.e.\ a strong preference for small $q$,
similar to what has been observed for G-dwarf systems (Duquennoy \& Major \cite{duquennoy91}).
The semimajor axis ($a$) is selected from a uniform distribution in $\log a$
with an upper cut-off at 3000~AU (Quist \& Lindegren \cite{quist00}). The
period distribution is afterwards generated by NBODY6 based on the modelling
by Kroupa (\cite{kroupa95a}, \cite{kroupa95b}) with minimum period 1~day, and binaries
merged if $a \le 10~R_{\odot}$.
The initial distribution of eccentricities $e$ is assumed to be thermal,
i.e.\ with a probability density function $2e$ (Kroupa \cite{kroupa95b}).

The only free model parameters are thus the total particle number and
the initial binary fraction (or multiplicity). Their determination
is discussed in Sect.~\ref{sec:fit}.

\subsection{Transformation to observables}
\label{sec:tran}

From NBODY6, the luminosity and temperature is obtained for each star.
These parameters are transformed to the observational plane $(B\!-\!V,M_V)$
using Kurucz's colour tables (e.g., Kurucz \cite{kurucz79} and
Buser \& Kurucz \cite{buser92}) for $\mbox{[Fe/H]}=0.10$.
Johnson's $V$ is used instead of the Hipparcos magnitude $Hp$, because of
the lack of adequate transformations for the latter. For binaries,
the combined colour and magnitude are calculated and plotted in order
to get results that are directly comparable with Hipparcos data.
In view of the very small interstellar reddening (Sect.~\ref{sec:bas}),
$E_{B\!-\!V}=0.0$ is assumed.

When comparing the simulated and observed HR diagrams it should be
borne in mind that the theoretical models and colour transformations
may produce non-negligible errors. Observed discrepancies for the
Hyades amount to some 0.05~mag in $B\!-\!V$ or 0.3~mag in $M_V$ in the
cool end of the main sequence (Castellani et al.\ \cite{castellani01}).
No (empiric) corrections for this effect have, however, been made.

In order to mimic the real Hyades cluster, as observed by Hipparcos,
the simulated present-day cluster is ``observed" from the same distance
as the real Hyades and given the same centroid velocity relative the Sun.
Small errors in the ``observed" $V$ magnitudes (standard deviation 0.0015~mag)
are introduced, and parallaxes and proper motions, including
observational errors, are generated following the same procedure as in
Lindegren et al.\ (\cite{lindegren00}). The simulated sample includes
all stars brighter than the completeness limit $V=7$, plus a random
selection of the fainter stars matching the real sample in the number
of stars per magnitude interval.
It is assumed that the Hyades stars in the Hipparcos Catalogue with $V > 7$ mag
are not subject to any selection effects, although it cannot be ruled out
due to a sometimes impenetrable selection procedure of Hipparcos objects
in open clusters (Mermilliod \& Turon \cite{mer89}).

Binaries receive different treatments depending on the magnitude
difference ($\Delta m$), period ($P$), and angular separation
($\rho$), in order to simulate how they were treated in the Hipparcos
data analysis (see Sect.~1.4.2 in the Hipparcos Catalogue).
Here, $\Delta m = M_{V2}-M_{V1}$, where subscripts ``1" and ``2" refer to 
the primary and secondary components.
For certain combinations of these parameters, Hipparcos effectively observed
the motion of the photocentre of the system. In the remaining cases the
centre of mass were observed. The former systems include
those with $P\simeq$ 0.1 to 20 years and $\rho \ge$ 10 mas,
or $P>$ 10 years and $\rho \le$ 100 mas; the short-period binaries
($P<$ few months), which may deviate from a single-star solution (the ``stochastic"
(X) solution), although they may have been detected as binaries by Hipparcos;
and systems with Hipparcos magnitude difference $\Delta Hp > 4$ mag.
For these systems, the component velocities
$\vec{v}_1$ and $\vec{v}_2$ are combined into a single velocity of
the photocentre,
\begin{equation}
    \label{eq:phv}
   \vec{v}_{\rm ph} = (1-\xi)\vec{v}_1+ \xi\vec{v}_2 \, ,
\end{equation}
where 
\begin{equation}
\xi=\frac{I_2}{I_1+I_2}=\frac{1}{1+10^{0.4\Delta m}}
\end{equation}
is the fractional intensity of the secondary. $I$ is the 
intensity for each component given by $I\propto 10^{-0.4M_V}$. For these
systems, a single proper motion is derived from $\vec{v}_{\rm ph}$.

The resulting simulated astrometric data are subject to exactly the same
maximum-likelihood estimation procedure as was used for the real cluster
(Lindegren et al.\ \cite{lindegren00}). In particular, astrometric radial
velocities and kinematically improved parallaxes are derived for the individual
stars or binaries. The error in the estimated parallaxes
is in the range 0.2--1.0~mas (an error of 0.5~mas corresponds to
approximately 1~pc in the cluster centre). The improved parallaxes are
used to compute distances from the cluster centre, which allow to count
the number of stars within a certain radius. Furthermore, for any
subsample of the stars, the velocity dispersion can be estimated from the
proper-motion residuals (Sect.~\ref{sec:kin}).

\begin{figure}
 \resizebox{8cm}{!}{\includegraphics*{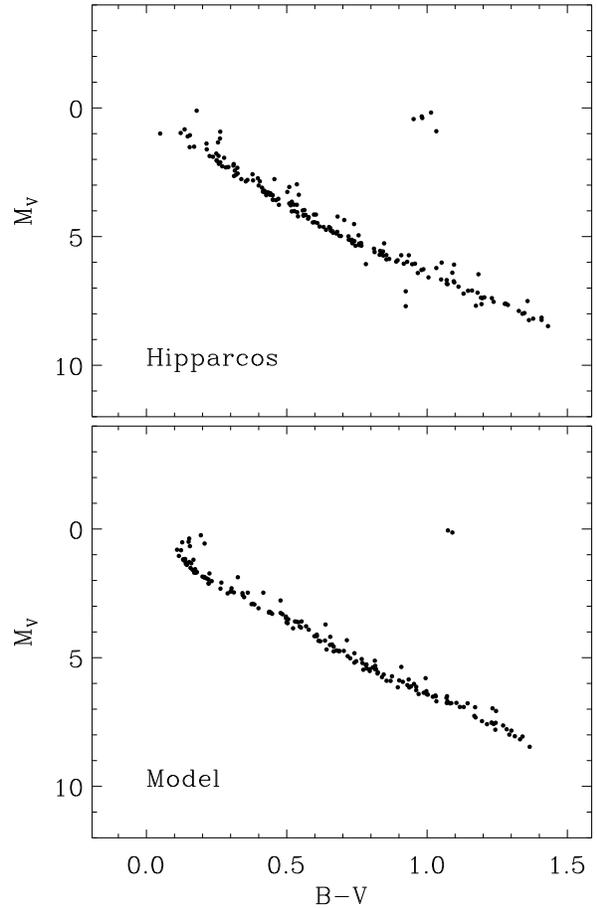}}
  \caption{The observational Hertzsprung-Russell diagram for the Hyades
cluster, based on Hipparcos data (top), and for one of several realisations
of the cluster model (bottom). In both cases the kinematically improved
parallaxes ($\widehat{\pi}$) are used. 
}
\label{fig:cmd}
\end{figure}

\subsection{Model fitting}
\label{sec:fit}

In order to tune the model parameters, it is necessary to make several
simulations for the same parameter values but using different
initialisations of the random number generator. The average of the
different random realisations is then compared with the observational
data, and the input parameters adjusted accordingly.
The quantities to be compared are the radial distribution of the stars,
their total number above a given magnitude limit, and binary statistics.
Also the number of giants (defined as $M_V<1$ and $B\!-\!V>0.5$) and
the number of single white dwarfs are used to constrain the model.

\begin{table*}[t]
  \caption[ ]{
The number of stars from Hipparcos ($N_{\rm HIP}$) and the mean of 20 model realisations
($N_{\rm model}$)
for certain constraints based on distance from the cluster centre $r$ and magnitude $V$.
Numbers after $\pm$ show the dispersion among the 20 realisations.
$r$ is calculated using either observed, estimated and true parallaxes. The latter
are, of course, not known for the real cluster.
Note that the term ``observed" in the table means both real
and simulated observations.
Giants are defined as stars with $B-V>0.5$ and $M_V<1$.
Note that white dwarfs are too faint to appear in the Hipparcos observations of the Hyades,
but since they are produced in the simulations, their number can be
compared with the minimum number from other observations.
}
\begin{tabular}{lrrrrrrrrr}
  \hline\noalign{\smallskip}
 & \multicolumn{2}{c}{observed parallaxes:} & & & \multicolumn{2}{c}{estimated parallaxes:} & & & true parallaxes: \\
   constraint                   & $N_{\rm HIP}$ & $N_{\rm model}$ & & & $N_{\rm HIP}$ & $N_{\rm model}$ & & & $N_{\rm model}$ \\
\noalign{\smallskip}
  \hline\noalign{\smallskip}
$ r \le 20$ pc                  & 173  & 166.7 $\pm$ 9.6 & & & 178  & 166.7 $\pm$ \phantom{0}9.7 & & & 167.6 $\pm$ 10.0  \\
$ r \le 10$ pc                  & 134  & 146.5 $\pm$ 9.3 & & & 143  & 149.5 $\pm$ 10.4 & & & 153.1 $\pm$  \phantom{0}9.9  \\
$ V \le 8$ mag ($r \le 10$ pc)  &  83  &  79.6 $\pm$ 6.1 & & & 88  &  81.3 $\pm$ \phantom{0}6.9 & & & 82.4 $\pm$  \phantom{0}6.5  \\
$ V \le 7$ mag ($r \le 10$ pc)  &  57  &  60.6 $\pm$ 6.1 & & & 58  &  61.8 $\pm$ \phantom{0}7.0 & & & 62.7 $\pm$  \phantom{0}6.4  \\
$ V \le 8$ mag ($r \le  3$ pc)  &  30  &  30.3 $\pm$ 5.8 & & & 38  &  37.0 $\pm$ \phantom{0}6.5 & & & 47.8 $\pm$  \phantom{0}8.4  \\
$ V \le 7$ mag ($r \le  3$ pc)  &  24  &  24.1 $\pm$ 4.4 & & & 29  &  30.2 $\pm$ \phantom{0}5.1 & & &  39.0 $\pm$  \phantom{0}6.3  \\
Giants ($r \le 20$ pc)          &   5  &   5.1 $\pm$ 2.0 & & &  5  &   5.1 $\pm$ \phantom{0}2.0 & & &    5.1 $\pm$  \phantom{0}2.0  \\
single white dwarfs  & $\ge 7\rlap{$\dagger$}$ &   & & &           &             & & &     8.5 $\pm$  \phantom{0}2.7  \\
\noalign{\smallskip} \hline
\noalign{\smallskip}
\end{tabular}\\
$\dagger$~from Reid (\cite{reid96}).
\label{tab:output}
\end{table*}

The finally adopted (protocluster) model comprises 200 single stars and
1200 binaries, i.e.\ an initial multiplicity of 0.86 companions per primary.
The total initial mass is $1100$--$1200~M_{\odot}$. This is slightly less
than previous estimates of $1200$--$1500~M_{\odot}$ (Reid \cite{reid93})
or $1300~M_{\odot}$ (Kroupa \cite{kroupa95c}). This smaller initial mass
was found necessary in order to correctly reproduce the number of 
observed stars with the given IMF. The true initial mass of the Hyades is
probably higher due to non-modelled mass loss (Sect.~\ref{sec:nmod}).
According to the simulations, the total current mass of the Hyades stars is
$\simeq$~460~$M_\odot$ with a tidal radius of $\simeq$~11~pc.
Observationally, Reid (\cite{reid92}) made the estimation 410--480$~M_{\odot}$
while Perryman et al.\ (\cite{perryman98}) estimated $400~M_{\odot}$
in their Hipparcos study of the cluster.

An example of the observational HR
diagram for one of the model realisations is shown in Fig.~\ref{fig:cmd},
together with the corresponding observed diagram for the Hyades cluster.
In addition to the standard deviation introduced in $V$, a standard deviation
of 0.01 in $B-V$ is also introduced in the model HR diagram to make the
colour distribution appear more realistic.
This standard deviation includes both observational errors and
the effects of peculiar stars, stellar rotation, etc.
 Apart from
the previously mentioned possible discrepancy in the cool end of the main
sequence, and the fact that the giant stars are too red in the simulations,
the general agreement is reasonable. The precise colours of the giants are,
however, irrelevant in the context of this study.

Table~\ref{tab:output} shows some statistics computed from this model, after
evolution to an age of 625~Myr and transformation to the observables, together
with the corresponding observed numbers. 
From Table~\ref{tab:output} it appears that the distribution of stars with radial distance
and apparent magnitude in the Hyades is well reproduced by the model cluster.
The number of stars decreases when we go from the constraints based on 
the true parallaxes to the constraints based on the estimated parallaxes,
and the number decreases even further when we use the observed parallaxes.
This is a result of observational errors affecting the parallaxes, and
mostly for the smallest sphere $r \le 3$.
In fact, the resemblence in the three columns is so good that it shows the modelling
of the errors are in accordance with reality.
The underabundance of stars in the models relative to the observations
in the range $10 < r < 20$~pc can be explained by an overestimation
of stars outside 10~pc by Perryman et al.\ (\cite{perryman98}). They
argued that using another centre of mass in the Hyades would lead
to fewer stars in the halo (Sect.~\ref{sec:bas}). 

It has been much more difficult to reproduce the observed binary statistics
(Table~\ref{tab:comp}). Bright binaries with high mass ratio
or small magnitude difference are underproduced. Even if every star in the
protocluster were assumed to be a binary (multiplicity 1.0), the model would
still predict too few binaries of these characteristics. 
The observed sample also has significantly more known short-period binaries
($P<10$~days) than obtained in the simulations. These discrepancies indicate
that the model distributions in mass ratio and/or semi-major axis would need
some adjustment. 
Alternatively, a higher initial mass leading to more binaries with the
required properties could be an explanation assuming non-modelled mass
loss of preferentially low mass stars.
However, the discrepancies are not dramatic and for the present
study it was preferred not to change the relevant code in NBODY6.

\begin{table}[b]
  \caption[ ]{
The number of Hyades binaries in the Hipparcos Catalogue ($N_{\rm HIP}$)
compared with the number from the mean of several random realisations of the
adopted cluster model ($N_{\rm model}$). The value after $\pm$ is the dispersion
around the mean value among the different realisations.}
\begin{tabular}{lrr}
\hline\noalign{\smallskip}
Constraint         & $N_{\rm HIP}$ &  $N_{\rm model}$   \\
\noalign{\smallskip}
\hline
\noalign{\smallskip}
\multicolumn{3}{c}{$r \le 20$~pc:} \\[3pt]
binaries, all       & $\ge 95$ & $137.8 \pm 8.8$ \\
binaries, $0.2<\rho<2\arcsec$, $\Delta m<2$ & $17$ & $13.7 \pm 3.7$ \\
binaries, $0.1<\rho<1\arcsec$, $q>0.4$ & $17$ & $10.3 \pm 3.2$ \\
binaries, $P<10$~days & $\ge 9$ & $3.6 \pm 1.6$ \\
\noalign{\smallskip}
\hline
\noalign{\medskip}
\end{tabular}
\label{tab:comp}
\end{table}

Since the initial multiplicity must be very
high to fit the observed binary statistics without being in contradiction
with the observed number of Hyades member stars, the degree of degeneracy 
between the two free input parameters (initial particle number and 
initial multiplicity) is small.

\begin{table*}[t]
\caption[ ]{
The number of stars ($N$) and observed velocity dispersion $\widehat{\sigma}_v$ in four
intervals of distance $r$ from the Hyades cluster centre, as estimated from the
proper-motion residuals in the Hipparcos and Tycho-2 catalogues and the kinematically
improved parallaxes. The Hy0r sample is the
``full" sample with 178 stars within 20~pc radius. Hy1r is the same sample but with all
known binaries removed. The Ty0r sample was created from Hy0r by replacing Hipparcos
proper motions with Tycho-2 ones, where available. Ty1r is the same sample but with all
known binaries removed. The last columns marked ``Model" give the average number of stars
and dispersions from 20 realisations of the adopted cluster model. $\widehat{\sigma}_v$
is the dispersion estimated as for the real cluster, while $\overline{\sigma}_v$ is the
``true" dispersion in the model, calculated from the three-dimensional peculiar velocities
relative the cluster centroid.
}
\begin{tabular}{@{\extracolsep{-2pt}}lrrrrrrrrrrrrrrrr}
\hline
\noalign{\smallskip}
  && \multicolumn{2}{c}{Hy0r} && \multicolumn{2}{c}{Hy1r} && \multicolumn{2}{c}{Ty0r} && \multicolumn{2}{c}{Ty1r} && \multicolumn{3}{c}{Model} \\[-5pt]
$g_{\rm lim}$ && \multicolumn{2}{c}{\hrulefill} && \multicolumn{2}{c}{\hrulefill} && \multicolumn{2}{c}{\hrulefill} && \multicolumn{2}{c}{\hrulefill} && \multicolumn{3}{c}{\hrulefill} \\
\noalign{\smallskip}
  && $N$ & \multicolumn{1}{c}{$\widehat{\sigma}_v$} && $N$ & \multicolumn{1}{c}{$\widehat{\sigma}_v$} && $N$ & \multicolumn{1}{c}{$\widehat{\sigma}_v$} && $N$ & \multicolumn{1}{c}{$\widehat{\sigma}_v$}
  && $\langle N \rangle$ & \multicolumn{1}{c}{$\langle\widehat{\sigma}_v\rangle$} & \multicolumn{1}{c}{$\langle\overline{\sigma}_v\rangle$} \\
\noalign{\smallskip}
\hline
\noalign{\smallskip}
\multicolumn{17}{c}{$r < 3$~pc:} \\[3pt]
$\infty$ && 55 & $0.70\pm 0.08$ && 20 & $0.32\pm 0.08$ && 60 & $0.39\pm 0.05$ && 20 & $0.22\pm 0.08$ && & &\\
15       && 51 & $0.42\pm 0.06$ && 21 & $0.30\pm 0.08$ && 57 & $0.30\pm 0.04$ && 21 & $0.20\pm 0.07$ && 54.8 & $0.45\pm 0.07$ & $0.33\pm 0.02$\\
10       && 45 & $0.21\pm 0.05$ && 20 & $0.26\pm 0.08$ && 52 & $0.24\pm 0.04$ && 18 & $0.22\pm 0.08$ && 48.7 & $0.32\pm 0.05$ & $0.32\pm 0.02$ \\
\noalign{\smallskip}
\hline
\noalign{\smallskip}
\multicolumn{17}{c}{$3 < r < 6$~pc:} \\[3pt]
$\infty$ && 56 & $0.83\pm 0.09$ && 30 & $0.34\pm 0.08$ && 58 & $0.39\pm 0.05$ && 28 & $0.28\pm 0.07$ && & &\\
15       && 53 & $0.47\pm 0.06$ && 27 & $0.33\pm 0.08$ && 58 & $0.39\pm 0.05$ && 27 & $0.28\pm 0.08$ && 50.6 & $0.44\pm 0.10$ & $0.28\pm 0.01$ \\
10       && 43 & $0.22\pm 0.05$ && 25 & $0.28\pm 0.08$ && 52 & $0.30\pm 0.05$ && 28 & $0.29\pm 0.07$ && 45.1 & $0.30\pm 0.06$ & $0.28\pm 0.01$ \\
\noalign{\smallskip}
\hline
\noalign{\smallskip}
\multicolumn{17}{c}{$6 < r < 10$~pc:} \\[3pt]
$\infty$ && 31 & $0.86\pm 0.13$ && 10 & $0.36\pm 0.12$ && 23 & $0.46\pm 0.09$ && 13 & $0.51\pm 0.12$ && & &\\
15       && 25 & $0.49\pm 0.09$ && 10 & $0.36\pm 0.12$ && 20 & $0.24\pm 0.07$ && 12 & $0.37\pm 0.11$ && 25.5 & $0.41\pm 0.10$ & $0.25\pm 0.02$ \\
10       && 20 & $0.29\pm 0.09$ && 11 & $0.34\pm 0.12$ && 20 & $0.18\pm 0.07$ &&  9 & $0.20\pm 0.10$ && 22.6 & $0.28\pm 0.07$ & $0.25\pm 0.02$ \\
\noalign{\smallskip}
\hline
\noalign{\smallskip}
\multicolumn{17}{c}{$10 < r < 20$~pc:} \\[3pt]
$\infty$ && 35 & $1.26\pm 0.16$ && 25 & $1.20\pm 0.18$ && 34 & $1.23\pm 0.16$ && 23 & $1.26\pm 0.19$ && & &\\
15       && 29 & $0.49\pm 0.09$ && 21 & $0.40\pm 0.10$ && 24 & $0.38\pm 0.08$ && 16 & $0.31\pm 0.10$ && 13.2 & $0.40\pm 0.13$ & $0.26\pm 0.02$ \\
10       && 24 & $0.25\pm 0.07$ && 18 & $0.33\pm 0.10$ && 24 & $0.33\pm 0.08$ && 17 & $0.29\pm 0.09$ && 11.6 & $0.26\pm 0.10$ & $0.27\pm 0.03$ \\
\noalign{\smallskip}
\hline
\end{tabular}
\label{tab:disp}
\end{table*}

The simulations could in principle be ``inverted" to derive an age,
by for instance stopping the modelling when the realisations appear
similar to observed structural or dynamical features in the Hyades.
But the non-modelled effects leading to mass loss during
the dynamical evolution will be a major uncertainty (Sect.~5).

\subsection{Observed kinematics versus simulated data}

\subsubsection{Dispersion versus cluster radius}

In a Plummer potential, the velocity dispersion decreases with cluster radius
according to Eq.~(\ref{eq:plum}). At some radius, however, the relation
is expected to break down when the stars have left the cluster
potential and become subject to the Galactic field. In the following this possible structure
is investigated.

The various observed samples (Hy0r, Hy1r, Ty0r, Ty1r), as well as the different
realisations of the adopted cluster model, are analysed by means of the
maximum-likelihood method mentioned in Sect.~\ref{sec:kin}. The samples
are divided according to distance ($r$) from the cluster centroid in order
to determine if there is a radial variation of the kinematics. The ranges in
$r$ have not been chosen at random: 3~pc is approximately the core radius
while 10~pc is approximately the tidal radius.
Table~\ref{tab:disp} summarises the results for the number $N$ of stars
(or systems) and the estimated velocity dispersion $\widehat{\sigma}_v$.

The analysis method includes the rejection procedure designed to ``clean" the
cluster membership described in Sect.~\ref{sec:kin} with the 
goodness-of-fit statistic $g_i$ calculated for each star.
For the model simulations, no results
are given for $g_{\rm lim}=\infty$ because of their sensitivity to run-away
stars. In the observed sample such cases were already removed by Perryman
et al.\ (\cite{perryman98}).
It should be noted that the cleaning process successively reduces the
estimated internal velocity dispersion, because the latter is based on the
proper-motion residuals, which are also reflected in $g_i$. This is most
clearly seen for the Hy0r sample at all radii, and for the other samples
at $r>10$~pc.
The reason that there seems to be more stars for e.g. Ty1r at
$g_{\rm lim}=15$ than $g_{\rm lim}=\infty$ for certain ranges in $r$ is that kinematically
improved parallaxes have been used to calculate the distance from cluster centre.
Since it is a different solution for each $g_{\rm lim}$, the kinematically
improved parallaxes may change slightly.

Kinematically, one cannot in general distinguish between actual non-member
stars and member stars with a deviating space motion. The most probable
reason for a member star not to follow the common space motion of the
cluster is that it is a binary in a non-modelled orbit.
As explained in Sect.~\ref{sec:bas}, this effect should be greater for the
samples based on the Hipparcos proper motions than when using the Tycho-2
data. Comparing the results for Hy0r and Ty0r as function of $g_{\rm lim}$
suggests that binaries are the main cause for deviating proper motions
out to $r \simeq 10$~pc, while for the greater radii they are partly caused
by actual non-members.

The last two columns in Table~\ref{tab:disp} show the estimated and
true dispersions from 20 realisations of the model. It appears that
$g_{\rm lim}=10$ yields a correct estimation of the dispersion,
while $g_{\rm lim}=15$ leads to an over-estimation of $\sigma_v$.
Using $g_{\rm lim}=10$, the cluster as a whole (inside 20~pc)
yields a dispersion of $0.23\pm 0.02$~km~s$^{-1}$, with no clear
dependence on $r$. The model cluster yields a slightly larger value
(0.30~km~s$^{-1}$) and shows a 20\% decrease from the centre outwards.
It should be noted that two of the 20 models yield estimated values
as small as the observations ($\le$~0.23~km~s$^{-1}$). The dispersions  
$\overline{\sigma}_v$ characterise the stars in the simulated Hyades
sample, and not the total number of stars in the cluster. Due to the
limiting magnitude, stars with masses less than 0.5--0.6~$M_\odot$
do not contribute to the velocity dispersions in the table, just
as with the observations. 

Madsen et al.\ (\cite{madsen01}) found some rather large radial variations
of the velocity dispersion in the Hyades, but could not conclude whether the
structure was real or not. From the present simulations it is concluded
that the observed structure is probably spurious: similar variations (of
either sign) can be seen in some of the model realisations, although
they are absent in the average of the realisations.

Hitherto in studies of open clusters, only in the Pleiades has an indication of
a relationship between $r$ and (the tangential component of)
$\sigma_v$ been found (van Leeuwen \cite{vanleeuwen83}).
In the globular cluster M15, however, a velocity dispersion decreasing
from the centre out to 7 arcmin and then increasing was found by
Drukier et al. (\cite{drukier98}). They interpreted it as an indication of
heating of the outer part of the cluster by the galactic tidal field. But how the
minimum at 7 arcmin was related to the tidal radius or other quantities remained unclear.
Heggie (\cite{heggie01}) argued that heating might be an incorrect interpretation since the
effect can also be seen in $N$-body simulations of star clusters moving under
influence of a steady tidal field (cf. Giersz \& Heggie \cite{giersz97}).
In the models here, the same trend is seen, with a minimum in the
$r-\sigma_v$ relation just inside 10~pc (the mean tidal radius of the models
is between 10 and 11~pc).

\subsubsection{Dispersion versus stellar mass}
\label{sec:dispmass}

Theoretically we should also expect a decreasing velocity dispersion
with higher mass, or correspondingly lower absolute magnitude,
due to equipartition of kinetic energy. This should in turn lead to
dynamical mass segregation, with the massive stars more concentrated
to the centre of the cluster.
This effect may have been seen in IC 2391 (Sagar \& Bhatt \cite{sagar89})
and Praesepe (Holland et al. \cite{holland00}).
Perryman et al.\ (\cite{perryman98}) found a clear mass segregation in
the Hyades from the number density of stars in various mass groups
as a function of distance from the centre. Direct searches by Lindegren
et al.\ (\cite{lindegren00}) and Madsen et al.\ (\cite{madsen01}) for
a relation between the observed velocity dispersion and mass (or absolute
magnitude), did however prove inconclusive.
Evidence of any equipartition of kinetic energy is best sought among the
stars in the core of the cluster (Inagaki \& Saslaw \cite{inagaki85}).
For the present study, a limiting radius of 3~pc is therefore used. This is 
approximately the core radius of the Hyades.

In the Hipparcos Catalogue, often only the common absolute magnitude for
a binary is available, and not the absolute magnitudes for both components.
Since it is the mass that is interesting, only the samples without
known binaries should be used, to ensure a reasonably unique
correspondence between absolute magnitude and mass.
In the simulated samples, 
binaries with a difference in absolute magnitude between the combined
absolute magnitude of the two components in the binary and the
the primary component of more than 0.1 mag have been removed. This
is the simplest way to simulate the hy1r sample.

The remaining stars with $r<3$~pc in the hy1r sample are separated in four intervals 
of absolute magnitude, with divisions at $M_V=2.1$, $3.4$, and $5.4$~mag,
approximately corresponding to the masses $1.8$, $1.4$, and $1.0 M_{\odot}$.
The estimated dispersions in these intervals are $0.17\pm 0.13$,
$0.20 \pm 0.11$, $0.24 \pm 0.11$~km~s$^{-1}$, and no solution for the
last interval. The uncertainties are too large to allow any firm conclusion,
although the expected trend is there. For comparison, the simulations gave
an average dispersion going from 0.28 to 0.36~km~s$^{-1}$ in the same
intervals.

\subsubsection{Other determinations of the dispersion}

Several studies of the velocity dispersion of the Hyades have been performed during the years.
In a detailed discussion by Gunn et al.\ (\cite{gunn88}), who performed a spectroscopic
investigation of the cluster, a mean dispersion of 0.23~km~s$^{-1}$ was derived from a Plummer
model. Their result agreed with the velocity dispersion obtained from
the most precise spectroscopic radial velocities
in their Hyades sample. However, it is important to note that the result of 0.23~km~s$^{-1}$ is
dependent on the estimated $M$ and $r_c$, where the mass is the major uncertainty.
Perryman et al.\ (\cite{perryman98}) also used a Plummer model and got 0.21~km~s$^{-1}$ for the
central velocity dispersion. Again this value was derived by estimating the mass
and the core radius.
Compared to this work the values are 50\% lower, but can be explained by
the uncertainty in the estimation of the masses.
Makarov et al.\ (\cite{makarov00}) used Tycho-2 proper motions to discuss the
velocity dispersion of the Hyades, and found the velocity dispersion to
be 0.32~km~s$^{-1}$ for the stars with the most precise proper motions.
If known spectroscopic binaries were removed, the velocity dispersion decreased
to 0.22~km~s$^{-1}$. The last value agrees well with the value obtained 
with Tycho-2 proper motions in Table~\ref{tab:disp}.

\section{Accuracy of astrometric radial velocities}
\label{sec:pred}

From the cluster simulations and subsequent application of the maximum-likelihood
method (Sect.~\ref{sec:kin}) the astrometric radial velocities are estimated
for the individual stars (or systems), $\widehat{v}_{{\rm r}i}$. Of course, the
true radial velocities $\overline{v}_{{\rm r}i}$ are also known directly from
the simulation. Thus the estimation errors
$\Delta_{ij}=\widehat{v}_{{\rm r}i}-\overline{v}_{{\rm r}i}$
are known. Here, index $j$ is used to distinguish the different realisations
of the cluster model. With $\langle\,\rangle_k$ denoting an average over index $k$,
the following statistics are computed:
\begin{equation}
  \Delta_j = \big\langle \Delta_{ij} \big\rangle_i
\end{equation}
is the ``cluster bias" in realisation $j$ (i.e., the common error for all stars in
the cluster);
\begin{equation}
  \epsilon_{\rm int} = \Big\langle (\Delta_{ij}-\Delta_j)^2 \Big\rangle_{ij}^{1/2}
\end{equation}
is the ``internal standard error" of the astrometric radial velocities (i.e., the
dispersion of the individual values around the cluster bias); and
\begin{equation}
  \epsilon_{\rm tot} = \Big\langle \Delta_{ij}^2 \Big\rangle_{ij}^{1/2}
\end{equation}
is the ``total standard error" of the astrometric radial velocities (i.e., including
the cluster bias). Clearly $\epsilon_{\rm int}$ is the relevant statistic for the
precision of \emph{relative} astrometric radial velocities within a given cluster,
while $\epsilon_{\rm tot}$ is relevant for the accuracy of \emph{absolute}
astrometric radial velocities. Both $\epsilon_{\rm int}$ and $\epsilon_{\rm tot}$
can be computed for various subsets depending on observable quantities such as
the goodness-of-fit measure $g_i$, radial distance $r$, and mass or absolute
magnitude. An interesting question is whether it is possible to observationally
define subsets with reduced $\epsilon_{\rm int}$ or $\epsilon_{\rm tot}$.

The results presented below are based on solutions using the rejection limit
$g_{\rm lim}=15$, although the results for $g_{\rm lim}=10$ are very similar.
Any conclusions from these simulations are also applicable to
the astrometric radial velocities published in Madsen et al.\ (\cite{madsen02}).

\subsection{Standard errors versus goodness-of-fit}

\begin{figure}
   \resizebox{\hsize}{!}{\includegraphics*{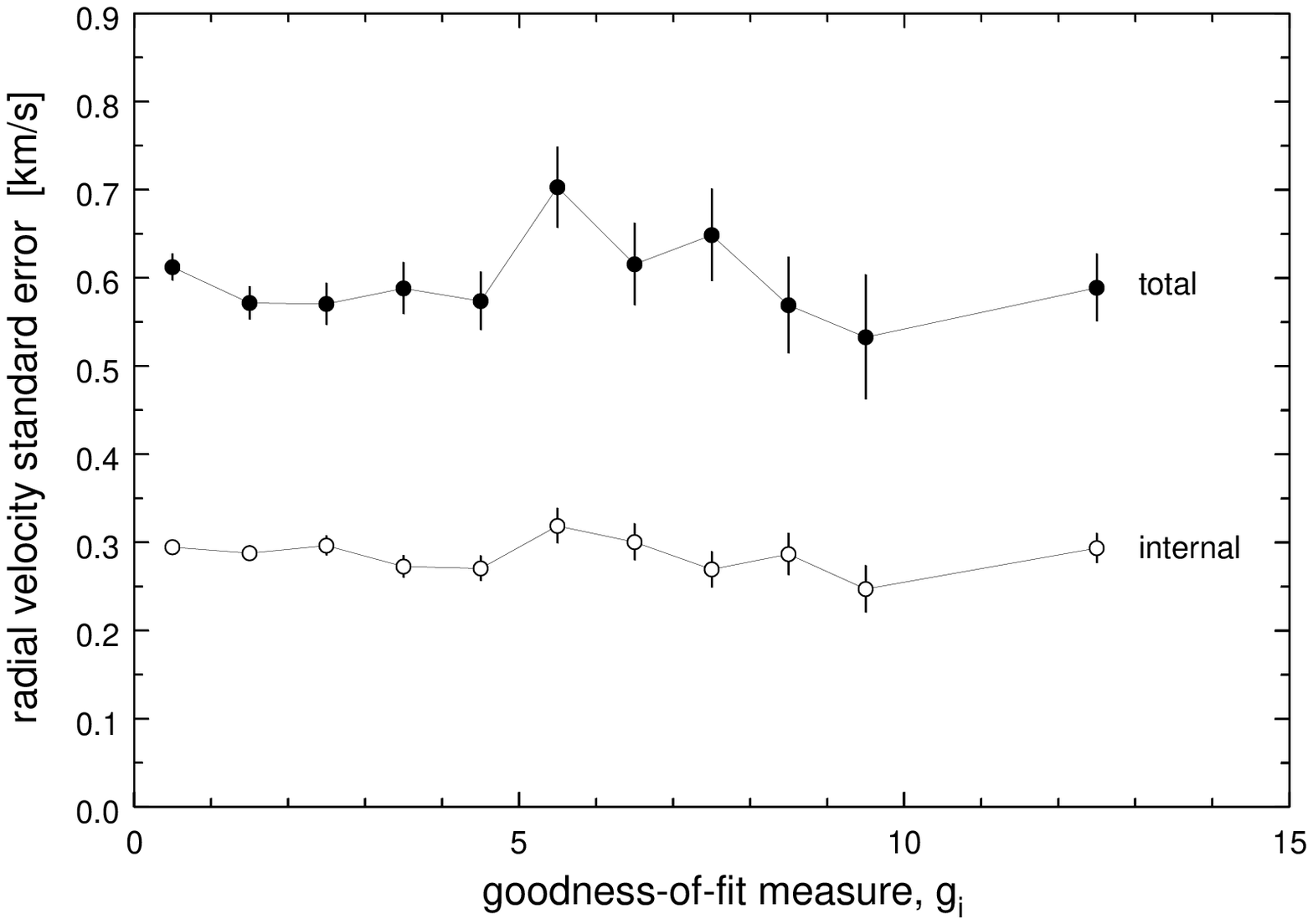}}
   \resizebox{\hsize}{!}{\includegraphics*{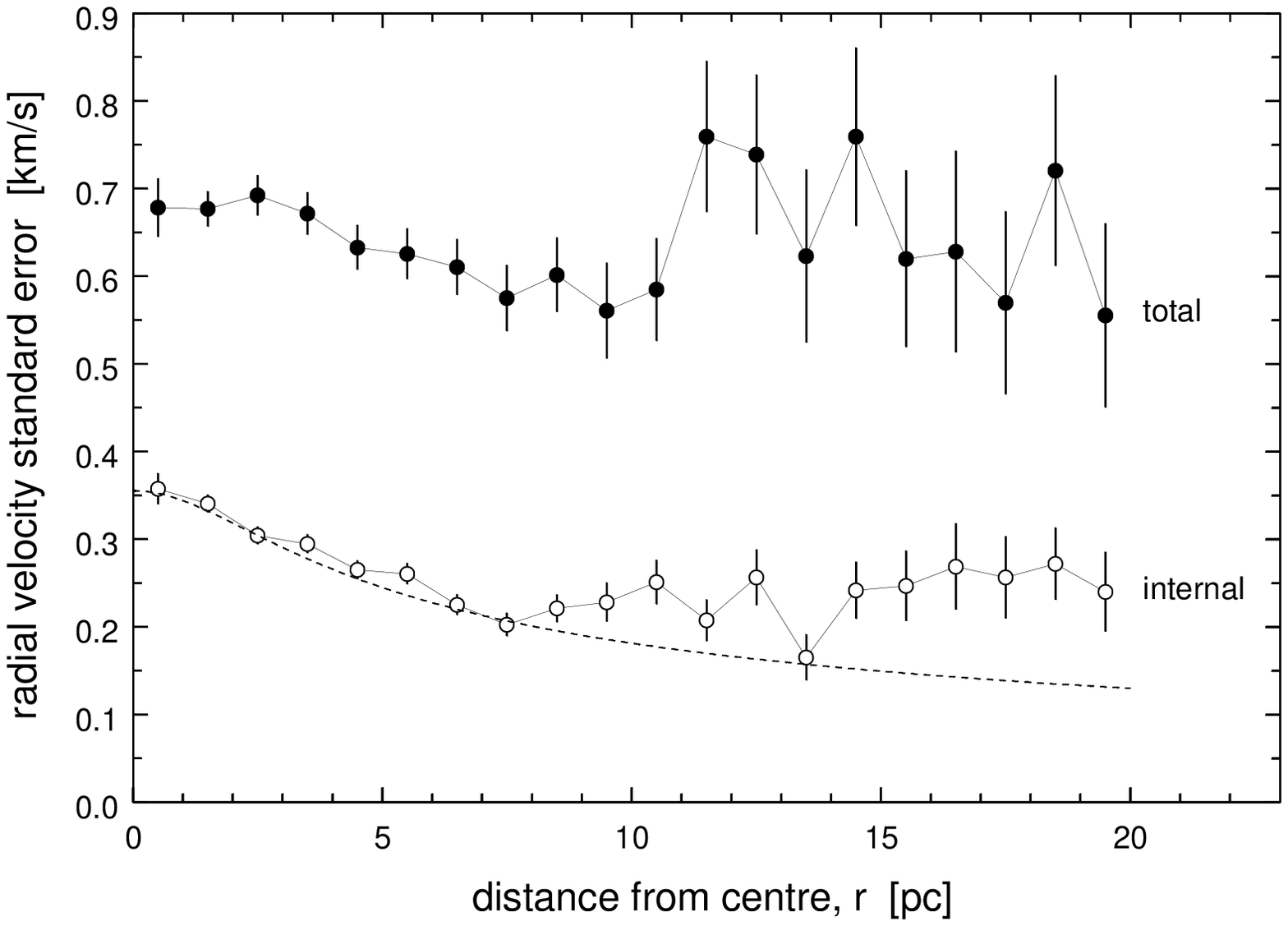}}
  \caption{Standard errors of the astrometric radial velocities as function
of the goodness-of-fit measure $g_i$ (top) and distance from the cluster
centre $r$ (bottom). Open circles show the internal standard errors
$\epsilon_{\rm int}$ (i.e., for the relative velocities within the cluster);
filled circles show the total standard errors $\epsilon_{\rm tot}$ (i.e., for
the absolute velocities). The dashed line is the expected relation
from the Plummer model.}
\label{fig2}
\end{figure}

In Fig.~\ref{fig2} (top) the internal and total standard errors of the astrometric
radial velocities are shown versus the goodness-of-fit $g_i$. The absence of
any significant trend shows that $g_i$ is not a useful criterion for selecting
``good" astrometric radial velocities. Even stars with $g_i>10$ are not worse
than the rest in terms of radial-velocity precision. This somewhat counter-intuitive
result can be understood if the line-of-sight component of the peculiar velocities
is statistically independent of the tangential component. This is obviously the
case for truly random motions, but one might expect that large proper-motion
errors caused by photocentric motion in binaries should be correlated with
large errors in the radial component.

\subsection{Standard errors versus radius}

The bottom part of Fig.~\ref{fig2} shows the internal and total standard errors
of the astrometric radial velocities versus the distance $r$ from the cluster
centre. In this case the standard errors clearly decrease from the centre out
to 7--8~pc radius, after which they seem to increase again.

The initial decrease (for $r<8$~pc) is roughly in agreement with the Plummer
model in Eq.~(\ref{eq:plum}) for $M\simeq 460~M_\odot$ and $r_c\simeq 2.7$~pc.

\subsection{Standard errors versus mass and absolute magnitude}

\begin{figure}
   \resizebox{\hsize}{!}{\includegraphics*{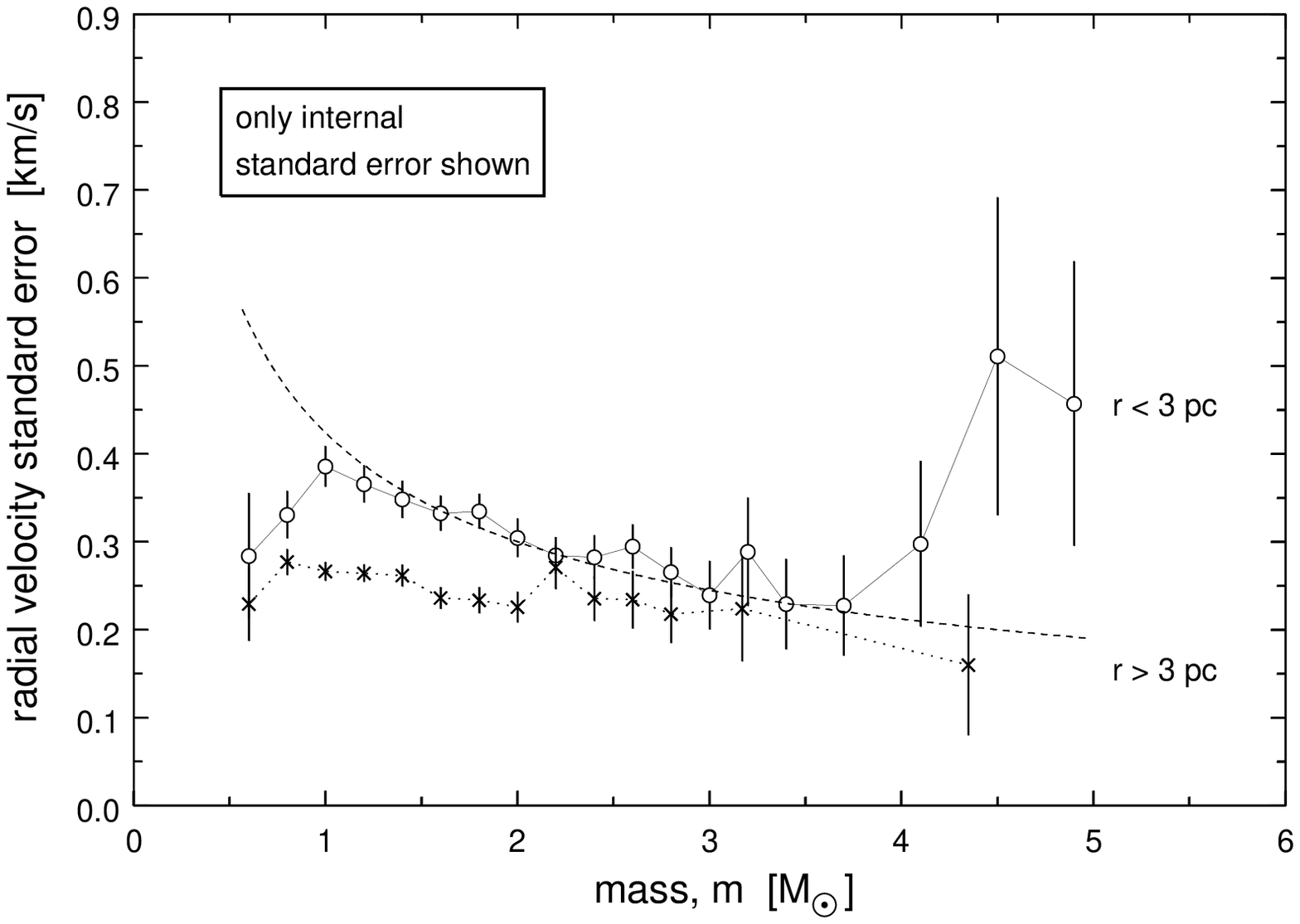}}
   \resizebox{\hsize}{!}{\includegraphics*{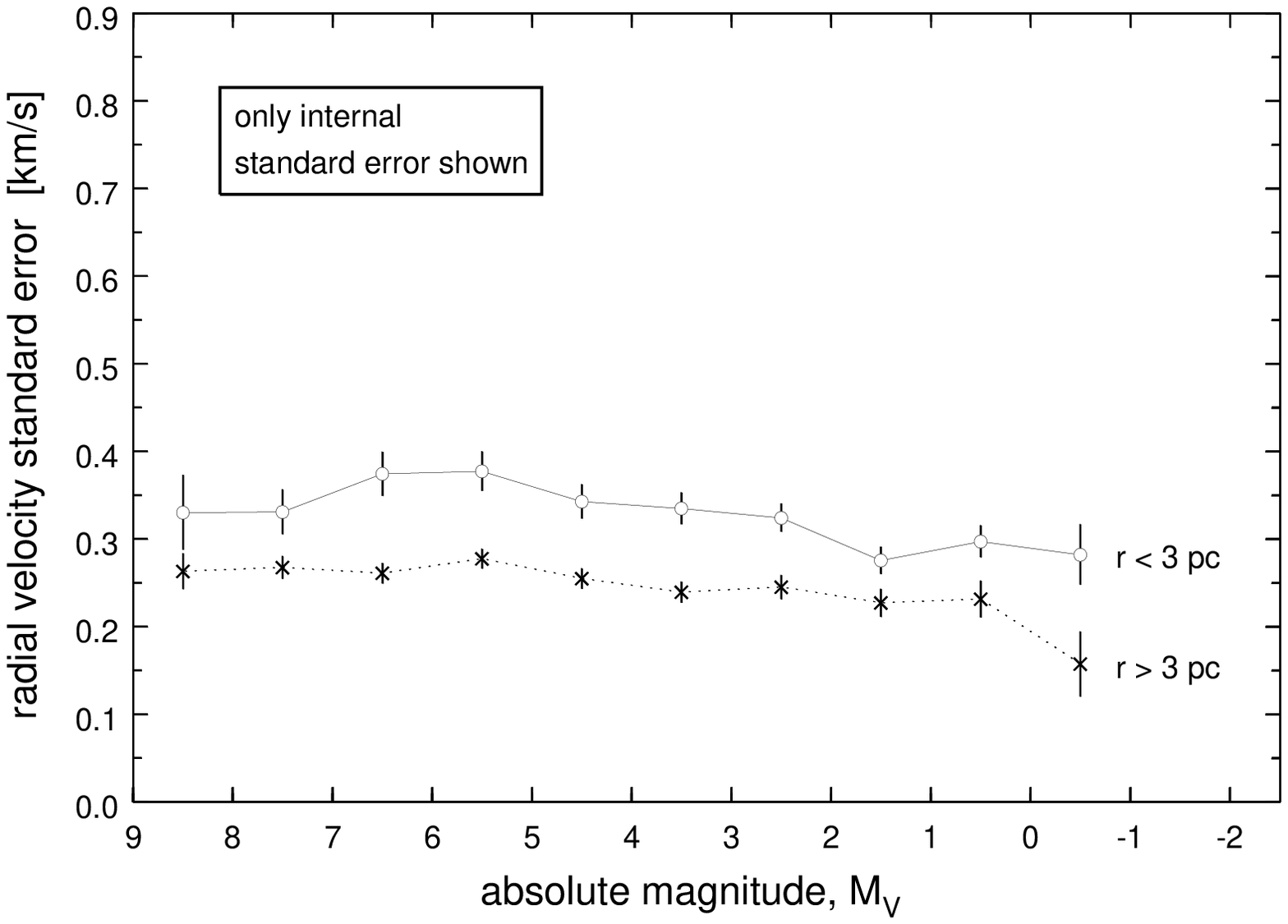}}
  \caption{The internal standard error of the astrometric radial velocities,
$\epsilon_{\rm int}$, as function of stellar mass (top) and absolute
magnitude (bottom). Circles refer to stars inside 3~pc of the cluster centre,
crosses to those outside 3~pc. For binaries, $m$ is the total mass of
the system and $M_V$ the total absolute magnitude.  The dashed line is
the curve $\sigma_v \propto M^{-1/2}$.}
\label{fig3}
\end{figure}

In Fig.~\ref{fig3}, the internal standard errors of the astrometric radial
velocities are plotted versus the true masses of the stars or systems (top)
and versus the absolute magnitudes (bottom). The sample is divided at
3~pc (see Sect.~\ref{sec:dispmass}). Inside 3~pc there is a
clear difference in the velocity dispersion between the highest
masses and 1 $M_{\odot}$, although not as much as for a full equipartition
of kinetic energy ($\sigma_v \propto M^{-1/2}$). The effect is much smaller outside
of 3~pc. The velocity dispersion
also seems to decline again for stars with masses less than 1 $M_{\odot}$.

The effect can still be seen when the dispersion is plotted versus absolute
magnitude instead of mass (Fig.~\ref{fig3}, bottom), although the trend is
less clear because of the many binary systems, for which there is no unique
correspondence between system mass and total luminosity.

Together with the results of the previous section we can conclude that the
practical minimum for the internal error of the astrometric radial velocities
in the Hyades is around 0.20~km~s$^{-1}$, which is achieved for stars at
an intermediate distance ($\simeq 2$--3 core radii) from the cluster centre.
At that distance there is little equipartition of kinetic energy, so it does
not matter much if more or less massive stars are selected.

\section{Non-modelled effects}
\label{sec:nmod}

The validity of the conclusions above depends critically on the realism of
the $N$-body simulations. A number of non-modelled effects, and their
possible impact on the results, are briefly considered below.

\emph{Time-dependent tidal field:}
When star clusters move through the galactic disk, they are subject to
tidal shocks, and shock heating from the bulge.
These effects are important to consider here since they increase
the random motion of the stars. For globular clusters it has been
found that tidal shocks accelerate significantly both core collapse
and evaporation (Gnedin et al.\ \cite{gnedin99}).

In the case of open clusters, Bergond et al.\ (\cite{bergond01}) estimated
that those with high-$z$ oscillations lose some 10--20\% of the mass 
integrated over the lifetime of the cluster,
mainly in low-mass stars, through disk-shocking.
The Hyades have a low vertical velocity ($W=6$~km~s$^{-1}$ relative to
the LSR), and therefore only oscillates with an amplitude of about
50 pc in $z$. Since this is small compared with the thickness of the
disk, the disk-crossings should not cause much additional heating.
The radial oscillations in the galactic plane, having an amplitude of
2~kpc, may be more important. The present $N$-body model assumes that
the cluster moves in a circular galactic orbit. Thus it cannot be excluded
that it underestimates the mass loss by perhaps some 5--10\% of the initial
mass. Preferentially, the lowest-mass stars leave the
cluster, forming tidal tails (Combes et al.\ \cite{combes99}). Although
this would slightly affect the estimation of the velocity dispersion,
it would have only a very small effect on the number of observed stars
of spectral type earlier than M0.

\emph{Molecular clouds:}
Terlevich (\cite{terlevich87}) studied open cluster $N$-body models with
initially 1000 particles and moving in a circular orbit at 10~kpc
from the galactic centre (i.e., assumptions comparable with this work).
She concluded
that the timescale for encounters with giant molecular clouds is of the
same order of magnitude as the present age of the Hyades. Since such
an encounter would probably be catastrophic, it can be assumed that the
Hyades have not been exposed to such a meeting. More abundant are
encounters with smaller interstellar clouds. They will not shorten the
lifetime of open clusters significantly but may contribute to the tidal heating
of the outer regions in a given cluster. Wielen (\cite{wielen75}) stated that
gravitational shocks due to interstellar clouds will produce a significant
flattening (up to 1:2) of the halo of the cluster perpendicular to the
galactic plane.
For the Hyades the flattening is 1:1.5 (Perryman et al.\ \cite{perryman98}).
Since the galactic tidal field is also contributing to the flattening, it is
doubtful if the Hyades have had any but minor interactions with interstellar clouds.

Perryman et al.\ (\cite{perryman98}) examined the possibility that the Hyades
recently experienced an encounter with a massive object causing a tidal
shear in the outer regions of the cluster, but excluded it based on the
impulsive approximation (Spitzer \cite{spitzer58}; Binney \& Tremaine \cite{binney87}).
Lindegren et al.\ (\cite{lindegren00}) included more velocity components in their
model to test for non-isotropic dilation, and concluded that if such an effect
existed it had to be higher than 0.01~km~s$^{-1}$~pc$^{-1}$ to be detected
with Hipparcos data. Effects from a tidal heating
are thus not detectable in the Hyades with current astrometric precision.

\emph{Brown dwarfs:}
Despite extensive searches, no single-star brown dwarf (BD) candidate has been
found in the Hyades (Reid \& Hawley \cite{reid99}; Gizis et al.\ \cite{gizis99};
Dobbie et al.\ \cite{dobbie02}). Reid \& Hawley (\cite{reid99}) found that the
lowest-mass Hyades candidate star (LH 0418+13) has a mass of 0.083~$M_{\odot}$,
placing it very close to the hydrogen-burning limit. The only promising candidate brown
dwarf in the Hyades is the unresolved companion in the short-period system RHy403
(Reid \& Mahoney \cite{reid00}). Of course, the faintness of these substellar
objects make them hard to observe, but still, the conclusion seems to be
that the number today is quite small.

Adams et al.\ (\cite{adams02}) performed extensive simulations with a
modified version of NBODY6 to model the brown dwarf population in
open clusters, and concluded that the effects of different
brown-dwarf populations were minimal, leaving the dynamics of the
cluster largely unchanged.

The IMF in the version of NBODY6 used here cannot produce brown dwarfs, so
this must be considered when defining the initial binary fraction.
The IMF for brown dwarfs, or substellar masses,
is very uncertain. Kroupa (\cite{kroupa01}) argues that a power-law value
of $\alpha=0.3\pm 0.7$ is the most reasonable.
Since stellar masses with $M<0.08M_{\odot}$ are not produced in the code,
one must represent the star--BD binary systems either as single stars or
by overproducing binaries with secondary components slightly above the
BD limit.
Thus an initial binary fraction of 86\% was assumed, which corresponds
approximately to unity if brown dwarfs had been included. Based on the
investigations of Adams et al. (\cite{adams02}), and considering that
Hipparcos did not observe stars less massive than M0 stars in the Hyades, the above
approximation should be sufficient for the present purpose.

\emph{Cluster rotation:}
Gunn et al.\ (\cite{gunn88}) did a comprehensive study of the rotation
of the Hyades, but had to conclude that it was at most of the same size
as their statistical error. Nonetheless they stated that their results
{\em suggested} a cluster rotation, but not higher than
0.015~km~s$^{-1}$~pc$^{-1}$.

Perryman et al.\ (\cite{perryman98}) did a thorough study of the
velocity residuals and concluded that they were consistent with
a non-rotating system and the given observational errors.
Lindegren et al.\ (\cite{lindegren00}) tested the Hyades for rotation
by assuming solid-body rotation parameters, but found that it was too small
to be detected, setting an upper limit of 0.01-0.02 km s$^{-1}$ pc$^{-1}$.
If this upper limit should equal the true rotation of the
Hyades, then the effect is non-negligible at 10~pc compared to the
internal error. But there seems to be nothing in the present
study suggesting such a rotation.

But probably the solid-body assumption is too simple. In the globular
cluster $\omega$ Centauri, Merritt et al. (\cite{merritt97}) found that
only at small radii could the rotation be approximated by a solid-body.
Beyond that the rotation falls off. Einsel \& Spurzem (\cite{einsel99})
did theoretical investigations on the influence of rotation on the dynamical
evolution of collisional stellar systems, that could explain
the findings by Merritt et al. (\cite{merritt97}). In fact, it seems
that only inside the half-mass radius is it reasonable
to talk about a solid-body rotation (cf. Kim et al. \cite{kim02}).

Although it is unlikely that the cloud in which the Hyades formed
had zero angular momentum, there currently exists no certain measure
of the rotation. In the model, it is instead assumed that the effects
are sufficiently small and can be ignored.

\emph{Expansion:}
During the evolution of a cluster parts of it expand and parts of
it contract. Under the assumption that the relative expansion rate
equals the inverse age of the cluster, Dravins et al.\ (\cite{dravins99b})
estimated that an isotropic expansion of the Hyades would lead to
a bias in the astrometric radial velocity of 0.07~km~s$^{-1}$ of
the centroid velocity. This is completely negligible and any
expansion effects have been ignored.

To summarise, it appears that none of these non-modelled effects would
affect the results very significantly. While the modelling of tidal fields
and brown dwarfs could be improved, the possible effect of cloud encounters
remains an uncertainty which cannot easily be included in the modelling
of a specific cluster such as the Hyades. Although NBODY6 allows encounters
with interstellar clouds, the option has not been used in the present study.

\section{Conclusions}
\label{sec:conclus}

A dynamical model of the Hyades cluster, based on $N$-body simulations
using the NBODY6 code, has been fitted to the astrometric information
available in the Hipparcos and Tycho-2 catalogues in order to
study the accuracy of astrometric radial velocities.
The number of stars as function of magnitude, their three-dimensional
distribution, and the distribution of proper motions have been
adequately reproduced by the model, as well as basic binary statistics.
No spectroscopic radial velocities have been used in the
present study (except for the initial membership determination by
Perryman et al.\ \cite{perryman98}) meaning that the results should
be directly comparable with the astrometrically determined radial
velocities of Hyades stars by Lindegren et al.\ (\cite{lindegren00})
and Madsen et al.\ (\cite{madsen02}). 

From the simulations it is concluded that the velocity dispersion of
the Hyades decreases from $\sigma_v\simeq 0.35$~km~s$^{-1}$ at the centre
of the cluster to nearly $0.2$~km~s$^{-1}$ at 7--8~pc from the centre.
Outside the tidal radius of 10--11~pc, the dispersion slightly increases
again. Compared with previous studies of the velocity dispersion in
the centre of the Hyades, the results here indicate a somewhat higher
value.

The internal velocity dispersion contributes to the random errors of the
astrometric radial velocities with the same magnitude. This is significantly
less than the $\sigma_v=0.49$~km~s$^{-1}$ estimated in Madsen et al.\
(\cite{madsen02}) directly from the Hipparcos observations. This discrepancy
can be understood with reference to Table~\ref{tab:disp} as an
\emph{overestimation} from the observed data when the less strict rejection
limit $g_{\rm lim}=15$ was used. Thus the previous estimate of the
internal standard error (due to the dispersion) can now be almost halved.

In fact, stars with an expected velocity dispersion as low as
0.20~km~s$^{-1}$ can be selected for studies that compare astrometric
and spectroscopic radial velocities in order to disclose astrophysical
phenomena causing spectroscopic line shifts. However, it should be
remembered that the total standard error, including the uncertainty
of the motion of the cluster centroid, is still of order
0.55--0.65~km~s$^{-1}$ (Fig.~\ref{fig2}, bottom), in agreement with
the previous estimate.

Attempts to see a radial dependence of the velocity dispersion
with Hipparcos and Tycho-2 astrometry have been inconclusive. The
observed relation is essentially flat for the most optimal sample.
Given the uncertainty of the estimated velocity dispersions when
the stars are divided into radial shells, this result is not surprising.
Similar examples can be found in the simulations. Only when the mean
relation is computed from several realisations of the cluster model
do the variations become clear.
In particular, it appears that the structure of dispersion/radius
relation reported by Madsen et al.\ (\cite{madsen01}) does not
reflect typical dynamical properties of the cluster, but could result
by chance or from some (unknown) mechanism related to the photocentric
motions of undetected binaries.

The fit has yielded an estimate of the initial cluster mass of
1100--1200~$M_\odot$ and of the initial multiplicity, which appears to
be very high (possibly near 100\%, if brown-dwarf companions are
included). The current cluster mass is estimated to be $\simeq$~460~$M_\odot$
with a tidal radius of $\simeq$~11~pc and a
mean velocity dispersion within $r<3$~pc of 0.32~km~s$^{-1}$.

Some of the differences between observations and simulations could be
due to some of the non-modelled features discussed in Sect.~\ref{sec:nmod},
which would lead to a higher initial particle number in the model and which might
also solve some of the discrepancies noted in the binary statistics.
The development of numerical tools such as NBODY6 to include
e.g.\ a time-dependent tidal field
would allow an improved realism of the Hyades model, and to study
the effect on the accuracy of astrometric radial velocities from
assumed negligible contributions to the velocity field with respect
to the Hipparcos precisions.

The method used to estimate astrometric radial velocities discussed
in Sect.~\ref{sec:kin} cannot eliminate of the error contribution
from the internal dynamics of the cluster, no matter how precise
the astrometry might be. The velocity dispersion therefore sets a fundamental limit
on the accuracy of astrometric radial velocities, and as a consequence the
results from the simulations presented here also apply to 
planned astrometric space missions such as GAIA (Perryman et al.\
\cite{perryman01}), even though it has been Hipparcos observations
of the Hyades that have been simulated.

The Hipparcos and Tycho-2 catalogues contain the best available
astrometry to study the internal velocity structure of the nearest
open cluster, the Hyades. To study it in greater detail, even better
astrometry is needed. The GAIA mission,
in combination with improved $N$-body simulations,
will make it possible to observe directly the internal velocity field of
the Hyades, and give us insight in the kinematics of the Hyades
in particular and open clusters in general.

\begin{acknowledgements}
I thank Sverre Aarseth for making NBODY6 freely available,
Tim Adams for helping me with the code, and Lennart Lindegren, Melvyn B.\ Davies,
and Dainis Dravins for useful comments and valuable suggestions.
\end{acknowledgements}

\end{document}